\documentclass[twocolumn,pra,superscriptaddress,showpacs,amsmath]{revtex4}

\usepackage{graphicx}

\begin{document}

\title
            {Resonant nonstationary amplification of
            polychromatic laser pulses and conical emission in an optically
            dense ensemble of neon metastable atoms}
\date{April 3, 2003}

\author{S. N. Bagayev}
\affiliation{Institute of Laser Physics, Siberian Branch of the Russian
Academy of Sciences, Lavrentyeva 13/3, 630090 Novosibirsk, Russia}
\author{V. S. Egorov}
\author{I. B. Mekhov}
\author{P. V. Moroshkin}
\author{I. A. Chekhonin}
\affiliation{St. Petersburg State University, Department of Optics,
Ulianovskaya 1, Petrodvorets, 198504 St. Petersburg, Russia}
\author{E. M. Davliatchine}
\author{E. Kindel}
\affiliation{Institut f\"ur Niedertemperatur-Plasmaphysik,
      Friedrich-Ludwig-Jahn-Str. 19, 17489 Greifswald, Germany}

\begin{abstract}
Experimental and numerical investigation of single-beam and pump-probe
interaction with a resonantly absorbing dense extended medium under strong
and weak field-matter coupling is presented. Significant probe beam
amplification and conical emission were observed. Under relatively weak
pumping and high medium density, when the condition of strong coupling
between field and resonant matter is fulfilled, the probe amplification
spectrum has a form of spectral doublet. Stronger pumping leads to the
appearance of a single peak of the probe beam amplification at the transition
frequency. The greater probe intensity results in an asymmetrical
transmission spectrum with amplification at the blue wing of the absorption
line and attenuation at the red one. Under high medium density, a broad band
of amplification appears. Theoretical model is based on the solution of the
Maxwell-Bloch equations for a two-level system. Different types of probe
transmission spectra obtained are attributed to complex dynamics of a
coherent medium response to broadband polychromatic radiation of a multimode
dye laser.
\end{abstract}


\maketitle

\section{Introduction}

We present a systematical experimental study of resonant interaction between
broadband pulsed-laser radiation and an optically dense extended two-level
medium. In our experiments, the pump-probe configuration was used and probe
beam amplification and/or attenuation were studied in detail. Single pump
beam propagation was investigated as well, and some off-axis radiation, which
is usually referred to as conical emission, was observed. A theoretical model
is based on the solution of the semiclassical Maxwell-Bloch equations taking
into account propagation effects of the electromagnetic field.

Investigation of resonant interactions in optically dense media is of
particular interest. In the case where the density of resonant particles is
high enough that a sufficient part of the external field can be coherently
absorbed and reemitted by the medium, a strongly coupled system of photons
and medium excitations, polaritons, is effectively created, and collective
phenomena play a key role in the interaction processes \cite{1}. Some
specific amplitude and phase (dispersive) characteristics that arise under
this regime lead to the appearance of coherent phenomena, which significantly
affect and enlarge the interaction picture. Particularly, many-atom vacuum
Rabi oscillations \cite{2,3} and optical ringing \cite{4} reflect the
coherent beating of two normal modes of the coupled field-matter systems,
which can be described as a splitting of polariton dispersion curve into two
branches. Such nonlinear effects as polariton parametric amplification in
semiconductor microcavities \cite{5,6,7} and a phenomenon of spectrum
condensation in atomic and molecular media \cite{8,9} were observed in
intracavity experiments under the large vacuum Rabi splitting.

The pump-probe experiments with gaseous resonantly absorbing media have been
reported in numerous publications. When the pump beam carrier frequency
$\omega_{\rm pump}$ is fixed and the probe beam frequency $\omega_{\rm
probe}$ is scanned near the frequency of an atomic transition $\omega_{0}$,
the probe-beam transmission has a form of the well-known Mollow-Boyd spectrum
\cite{10,11,12}. Its spectral features are determined by a value of the
generalized Rabi frequency of the strong field applied. Although the theory
\cite{10,11,12} is developed  for the steady-state regime, it describes the
pulsed-laser experiments \cite{13,14} as well as experiments with cw lasers
\cite{15}. Recently, the steady-state theory was generalized for
nonstationary processes using numerical solution of the Maxwell-Bloch
equations \cite{16}. In all of these experiments, the generalized Rabi
frequency of the pump beam is much greater than the spectral widths of the
atomic transition, $\gamma_{2}$, and both laser pulses $\gamma_{\rm pump}$
and $\gamma_{\rm probe}$. Besides that, a medium coherently absorbs and
reemits a small portion of the laser pulse energy, hence the field of the
medium response can be neglected with respect to the strong external field.
In this case, the theory of light-matter interaction is reduced to the
consideration of a single atom driven by the strong electromagnetic field.

Conical emission (CE) in gaseous atomic media was discovered by Grischkowsky
in 1970 \cite{17} and investigated in detail by many researchers. A
comprehensive review can be found in Ref.~\cite{18}. CE is usually observed
when intense laser radiation (cw or pulsed), detuned to the blue side from an
atomic resonance, propagates in a dense resonant medium. The spectrum of CE
is usually much broader than $\gamma_{\rm pump}$ and shifted to the red side
of an absorption line. Despite all efforts, only CE under cw pumping has
received a complete theoretical explanation \cite{19}. Moreover, several
different types of CE were observed in some experiments \cite{20,21},
suggesting that several possible mechanisms of CE generation exist. Influence
of collective effects on the conical emission was discussed in
Refs.~\cite{14,22}.

In cw experiments, spectral width of the laser radiation is usually much less
than both homogeneous $\gamma_{2}$ and inhomogeneous (Doppler) $\gamma_{D}$
widths of an atomic transition. In all pulsed experiments discussed above,
laser radiation has a form of smooth bell-shaped pulses with spectral width,
determined by the pulse duration such that $\gamma_{\rm pump},\gamma_{\rm
probe}> \gamma_{2}$ and $\gamma_{\rm pump},\gamma_{\rm probe}$ are of the
same order of magnitude as $\gamma_{D}$. In the present work, we consider
quite a different situation: the spectral width of laser pulses is determined
by the number of longitudinal modes (more than 100) of a multimode dye laser
and is much greater than the inverse pulse duration and both $\gamma_{2}$ and
$\gamma_{D}$. This enables us to study interaction of radiation with both red
and blue polariton branches simultaneously. The envelope of such a broadband
polychromatic pulse consists of a great number of very short irregular peaks.
Detunings of the pump and probe pulses from the transition frequency, defined
as $\Delta_{0}=\omega_{\rm pump}-\omega_{0}$, $\delta_{0}=\omega_{\rm
probe}-\omega_{0}$, are much smaller than the spectral widths of both lasers,
$\gamma_{\rm pump}$ and $\gamma_{\rm probe}$ (here $\omega_{\rm pump}$ and
$\omega_{\rm probe}$ correspond to mean frequencies of the pump and probe
spectra). The pump beam intensity $J_{\rm pump}$ was scanned in the range
that corresponds to the Rabi frequency $\Omega < \gamma_{\rm pump}$. The
relation between values of $\gamma_{\rm pump}$, $\gamma_{D}$, and $\Omega$
considered in the present paper characterizes a system, which up to now was
not investigated systematically. Resonant interaction of atomic media and
broadband pulses was considered only in some works on self-induced
transparency (SIT) in the sharp line limit \cite{23,24} (see also
Ref.~\cite{38}) and in intracavity experiments, where the phenomenon of
self-frequency-locking, or spectrum condensation, was studied \cite{8,9}.

The paper is organized as follows. In Sec.~II, our experimental setup is
described. In Sec.~III, the experimental results are presented, being divided
into three parts: probe-beam transmission under extremely broadband pump,
probe-beam transmission under relatively narrow-band pump, and conical
emission measurements. Section~IV is devoted to the theoretical model,
numerical simulation, and discussion of the experimental data. Main results
are summarized in Sec.~V.

\section{Experimental setup}

\begin{figure*}
\includegraphics{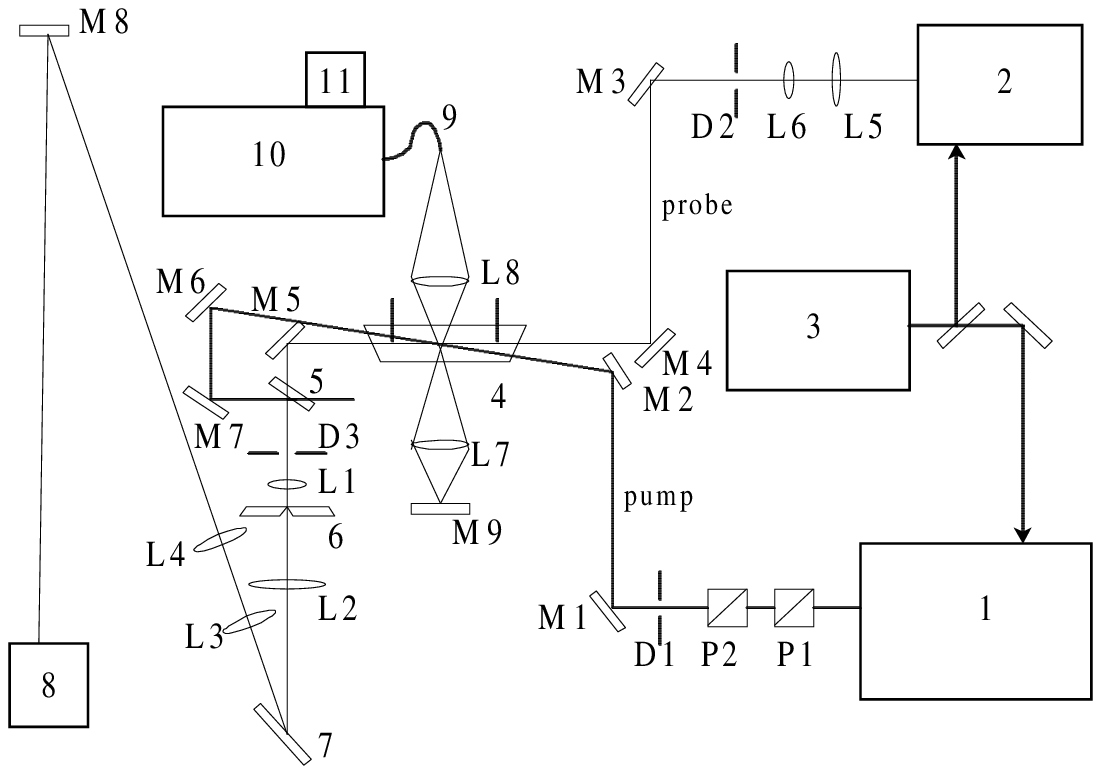}
\caption{\label{fig1}Scheme of the experimental setup: 1, 2, dye lasers; 3,
Nd:YAG laser; 4, discharge tube; 5, beam splitter; 6, slit; 7, diffraction
grating (2400 grooves/mm); 8, optical multichannel analyzer (OMA); 9, fiber;
10, spectrograph; 11, OMA; M1 - M9, mirrors; L1 - L8, lenses; D1 - D3,
diaphragms; P1, P2, polarizers.}
\end{figure*}

The scheme of our experimental setup is presented in Fig.~\ref{fig1}. We used
two pulsed multimode dye lasers (1, 2) pumped by the same Nd:YAG (where YAG
stands for yttrium aluminium garnet) laser (3) for the pump-probe
experiments. The pulse duration of both dye lasers was about 7 ns, the
probe-beam spectral width was about 1 nm. The dye laser (1) producing the
pump beam has two operation modes with different spectral widths:
``broadband'' mode $\gamma_{\rm pump}/2\pi \approx 300$ GHz (with a prism
inside the cavity) and ``narrowband'' mode $\gamma_{\rm pump}/2\pi \approx
10$ GHz (with a diffraction grating inside the cavity). The pump beam
intensity $J_{\rm pump}$ was scanned in a range of $10^4$ - $10^6$ W/cm$^2$
(measured pulse energies 1-100 $\mu$J) by rotation of a polarizer P2. The
pump and probe beams were intersected at the angle of about 1$^\circ$ inside
a discharge tube (4) containing a resonant medium. The probe beam spectrum
was analyzed by means of a high-resolution spectrograph constructed on the
optical table. The spectral resolution was about 12 GHz. The spectra were
recorded by a charge-coupled device (CCD) camera (8) with 100 ns exposure
time. Because of great fluctuations of the multimode dye-laser intensity, we
averaged recorded probe beam spectrum over 500 pulses. To control wavelength
tuning of the dye laser (1), part of its radiation was directed to the same
high-resolution spectrograph by a beam splitter (5) and a system of mirrors.

The wavelengths of both dye lasers were tuned to a transition of neon 1s$_5$
- 2p$_9$ in Paschen notation ($\lambda$=640.2 nm) with a metastable lower
state. This is the strongest red line of neon spectrum. A pulsed neon
discharge was used to produce sufficient amount of metastable neon atoms. The
discharge tube has a diameter of 10 mm and the length $L=12$ cm. The
amplitude of the discharge current was $I=4.0$ - 4.5 A, pulse duration $t_p=
80$ - 100 $\mu$s, neon pressure $P=9$ - 14 Torr. Under these conditions in
the discharge afterglow, the metastable atom density $n_0$ reaches great
values up to $10^{13}$ cm$^{-3}$. The optical density was scanned in a range
of $\alpha_{0}L \approx 20$ - 200 by changing the time delay $\tau_{d}$
between the discharge current and the laser pulse. Here $\alpha_{0}$ is the
absorption coefficient at the center of Doppler-broadened absorption line.
Experiment was also performed with a dc discharge in neon in the same
discharge tube under the conditions of $I=2$ - 50 mA and $P=0.5$ - 3.5 Torr.
In a dc discharge, metastable atom density was about $10^{12}$ cm$^{-3}$. We
controlled this value by measuring the resonant absorption across the
discharge tube. A well-known technique with a mirror behind the discharge
tube was used (mirror M9 in Fig.~\ref{fig1}). In this configuration the same
discharge tube serves as an absorber and a light source. A spectrograph (10)
with a CCD camera (11) was used for these measurements.

\section{Experimental results}

\subsection{Pump-probe experiment with broadband pump}

\begin{figure}
\includegraphics{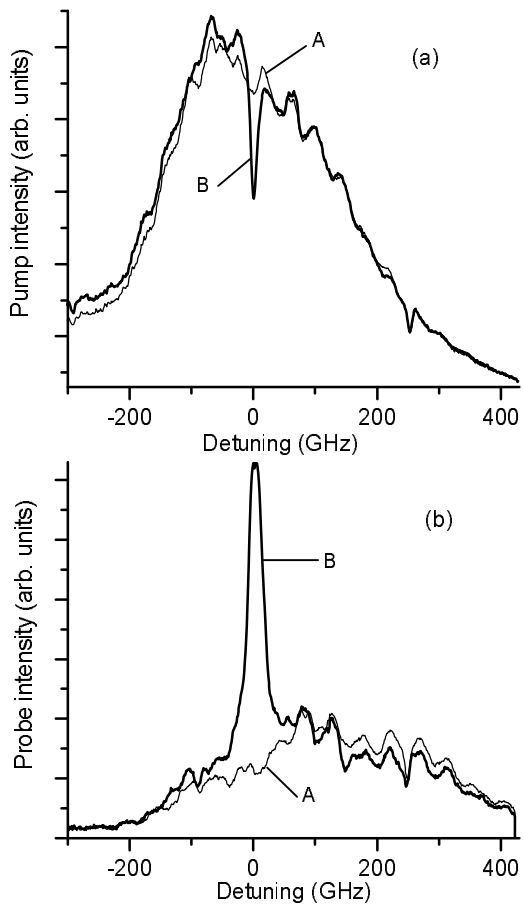}
\caption{\label{fig2}Dye-laser spectra: (a) the pump beam in a broadband
mode; (b) the probe beam; dc discharge, $P=0.9$ Torr, $I=20$ mA,
$\lambda=640.2$ nm, $n_0=10^{12}$ cm$^{-3}$, $\alpha_0 L=20$; A, original dye
laser spectrum; B, the spectrum after interaction with the medium.}
\end{figure}

\begin{figure}
\includegraphics{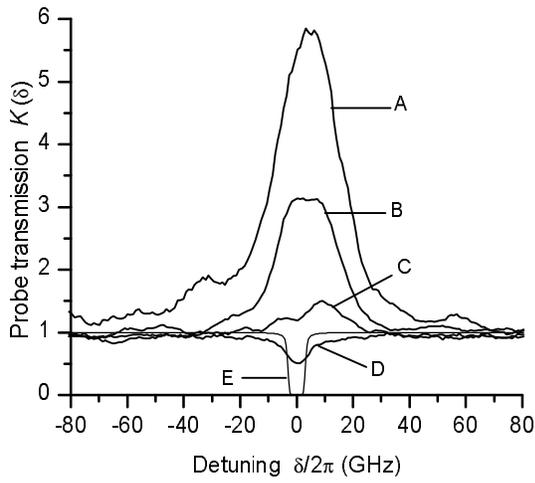}
\caption{\label{fig3}Probe-beam transmission spectra at different pump
powers; dc discharge, $P=1.2$ Torr, $I=10$ mA, $\lambda=640.2$ nm,
$n_0=10^{12}$ cm$^{-3}$, $\alpha_0 L=20$, broadband pump; A, $W=76$ $\mu$J;
B, $W=60$ $\mu$J; C, $W=30$ $\mu$J; D, $W=0$; E, calculated classical
absorption-line contour.}
\end{figure}

\begin{figure}
\includegraphics{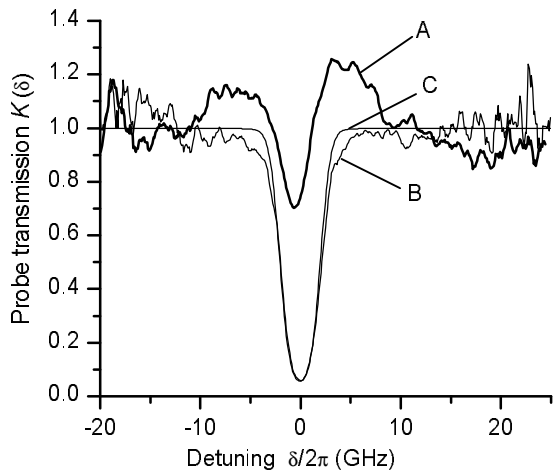}
\caption{\label{fig4}Probe-beam transmission spectra recorded with a
Fabry-Perot interferometer; dc discharge, $P=1$ Torr, $I=50$ mA,
$\lambda=588.2$ nm, $n_0=6\times 10^{11}$ cm$^{-3}$, $\alpha_0 L=2$, $W
\approx 1$ $\mu$J; A, the pump beam on;  B, the pump beam off;  C, calculated
classical absorption-line contour.}
\end{figure}

\begin{figure}
\includegraphics{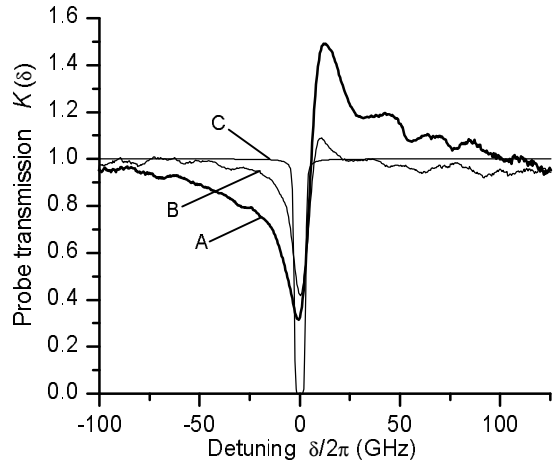}
\caption{\label{fig5}Probe-beam transmission spectra in the case of strong
probe beam; dc discharge, $P=1.2$ Torr, $I=10$ mA, $\lambda=640.2$ nm,
$n_0=10^{12}$ cm$^{-3}$, $\alpha_0 L=20$, $W=26$ $\mu$J, broadband pump;  A,
the pump beam on;  B, the pump beam off; C, calculated classical
absorption-line contour.}
\end{figure}

\begin{figure}
\includegraphics{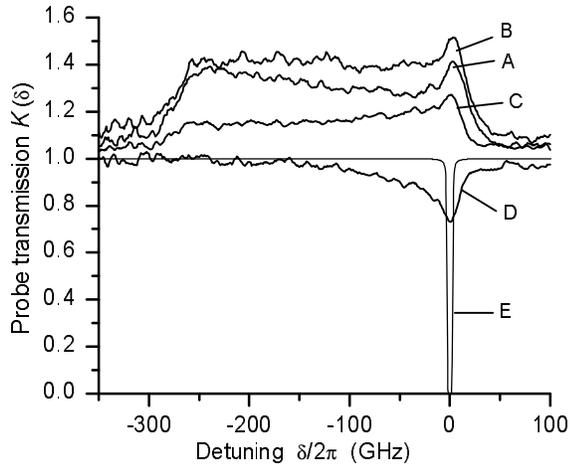}
\caption{\label{fig6}Probe-beam transmission spectra at different time
delays; pulsed discharge afterglow, $P=9.0$ Torr, $I=4.0$ A, $t_p=100$
$\mu$s, $\lambda=640.2$ nm, $n_0 \approx 10^{13}$ cm$^{-3}$, $\alpha_0 L
\approx 200$, $W=76$ $\mu$J, broadband pump; A, $\tau_d=30$ $\mu$s; B,
$\tau_d=80$ $\mu$s; C, $\tau_d=180$ $\mu$s; D, $\tau_d=30$ $\mu$s with pump
beam off; E, calculated classical absorption-line contour.}
\end{figure}

\begin{figure}
\includegraphics{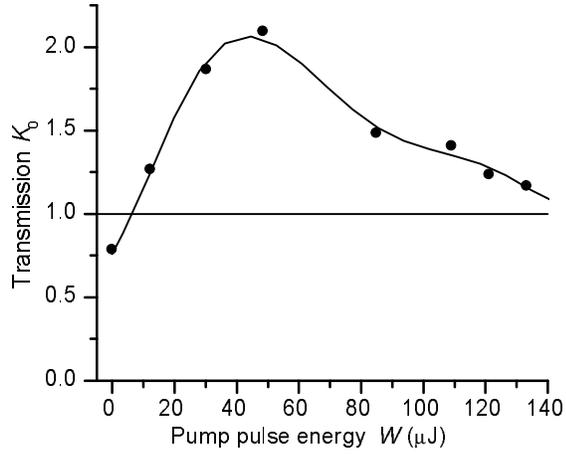}
\caption{\label{fig7}Dependence of the transmission at the resonance
frequency $K_0$ on the pump pulse energy $W$;  dc discharge, $P=1.2$ Torr,
$I=10$ mA, $\lambda=640.2$ nm, $n_0 =10^{12}$ cm$^{-3}$, $\alpha_0 L=20$,
broadband pump, transmission spectrum belongs to the first type.}
\end{figure}

The results of our pump-probe experiments with broadband pump beam are
presented in Figs.~\ref{fig2} - \ref{fig7}. In Fig.~\ref{fig2}(a), typical
spectra of the pump beam with discharge switched off (curve A) and on (curve
B) are plotted. The absorption line contour is partially resolved and shows
the position of the atomic resonance. Here the pump pulse energy $W$ is about
30 $\mu$J. At greater power there is no absorption because of the medium
saturation. Figure~\ref{fig2}(b) shows the probe beam spectra under the same
conditions with pump beam switched on (the pump pulse energy $W \approx 80$
$\mu$J).

We observed significant (up to six times) amplification of some spectral
components of the probe beam. In Figs.~\ref{fig3} - \ref{fig6}, the most
characteristic probe transmission spectra $K(\delta)$ are presented, where
$K(\delta)=J(\delta)/J^{0}(\delta)$, $\delta = \omega -\omega_0$ is the
detuning from the atomic resonance, $J(\delta)$ is the probe beam spectrum
after its interaction with the medium, and $J^{0}(\delta)$ is the original
probe beam spectrum. Systematical investigation of these spectra has shown
that there exist four amplification modes with different characteristic
shapes of the amplification spectrum: (i) spectrum with a single peak at the
resonance frequency (curve A in Fig.~\ref{fig3}); (ii) doublet centered at
the resonance frequency with partially saturated absorption line (curve C in
Fig.~\ref{fig3}, Fig.~\ref{fig4}); (iii) spectrum of dispersionlike shape
(Fig.~\ref{fig5}); and (iv) spectrum with a resonant peak and a broad
``pedestal'' (Fig.~\ref{fig6}). Four separate series of measurements are
presented below to clarify how the shape of the spectrum depends on the main
parameters of our experiment: $J_{\rm pump}$, $J_{\rm probe}$, $n_0$, and
$\Delta_{0}$. The latter dependence was performed under the narrowband
pumping and is presented in Sec. III B.

\subsubsection{Scanning of the pump intensity}

Transmission spectra of the probe beam at different pump powers under fixed
$J_{\rm probe}$ are presented in Fig.~\ref{fig3}. Curve A in Fig.~\ref{fig3}
corresponds to the pair of spectra in Fig.~\ref{fig2}(b) and represents
amplification spectrum of the first type, with a single resonant peak. It
appeared under significant pump beam intensities ($W > 30$ $\mu$J). The
dependence of the transmission at the resonance frequency $K_0$ on the pump
pulse energy $W$ is shown in Fig.~\ref{fig7}.

Decrease in the pump beam power below $W = 30$ $\mu$J leads to the splitting
of this peak into a doublet, i.e., spectrum of the second type, as it is
shown by curves B and C in Fig.~\ref{fig3}. Spectra of the second type were
investigated in detail by means of a scanning Fabry-Perot interferometer with
500 MHz spectral resolution (cf. Refs.~\cite{25,26}). For this experiment
both the pump and probe beams were obtained from the same dye laser with the
spectral width $\gamma_{\rm pump}/2\pi \approx 30$ GHz. Different transition
with the same metastable lower state was used ($\lambda$=588.2 nm). One can
find a more detailed description of the experimental setup and the results
obtained in Refs.~\cite{25,26}. Typical transmission spectra are presented in
Fig.~\ref{fig4}. As it is shown, the components of the doublet are
symmetrically detuned from the resonance frequency. There is significant
absorption near the resonance. Both components of the doublet decrease, with
pump beam power decreasing and completely disappearing when light-matter
interaction becomes linear.

\subsubsection{Strong probe beam}

Increase in the probe beam power results in transmission spectra of the third
type (Fig.~\ref{fig5}). Some asymmetry in transmission spectrum appears under
greater probe beam power even without the pump beam. This is clearly seen in
Fig.~\ref{fig5} (curve B), where the single probe beam transmission is shown.
For the case of the weak probe beam, there is no such asymmetry, as one can
see in Fig.~\ref{fig3} (curve D) and Fig.~\ref{fig4} (curve B). With the pump
switched on, this asymmetry becomes much more pronounced. There is strong
amplification in the blue wing of the absorption line and attenuation in the
red wing. Antisymmetrical spectrum, shown by curve A in Fig.~\ref{fig5}, was
observed under moderate pumping ($W = 20 - 30$ $\mu$J). Decrease in the pump
beam power under fixed probe power leads to the decrease in the amplification
coefficient at blue detunings, while at higher pump intensities the
amplification increases. In the latter case the probe-beam energy integrated
over the whole spectrum is amplified, which evidenced the existence of energy
transfer from the pump beam.

\subsubsection{Increase of the optical density}

Spectra of the first, second and third type were observed in a broad range of
metastable atom densities $n_0$: from $5$x$10^{11}$ cm$^{-3}$ (in a dc
discharge), up to $10^{13}$ cm$^{-3}$ (in a pulsed discharge afterglow).
Increase in the $n_0$ value leads to higher efficiency of the probe
amplification in both the first and second amplification regimes. In the case
of the strong probe beam (third regime), evolution of the transmission
spectrum is more complicated. The increase in $n_0$ resulted in (i)
broadening of the dispersionlike feature, (ii) increase in the absorption at
the resonance frequency, and (iii) increase in the amplification at the wing
of the absorption line.

Moreover, at maximum atomic density, another, fourth type of the
amplification spectrum was observed (Fig.~\ref{fig6}). In the pulsed
discharge experiment, the metastable atom density rapidly increases in the
afterglow, when the discharge current is switched off, reaches its maximum
value (about $10^{13}$ cm$^{-3}$), and then slowly decreases (the time
interval between the discharge pulses was much greater than the
characteristic decay time of $n_0$). Curves A - C in Fig.~\ref{fig6} show how
the spectrum of the fourth type varies with the time delay between the
discharge and laser pulses. Amplification reaches its maximum at the delay of
80 $\mu$s, when $n_0$ has a maximum. The transmission spectrum of this type
is a superposition of the resonant peak, described above as the first-type
spectrum, and the broadband amplification at the red side of the absorption
line. The spectral width of this band is extremely large, up to 350 GHz.

\subsection{Pump-probe experiment with narrowband pump}

\begin{figure}
\includegraphics{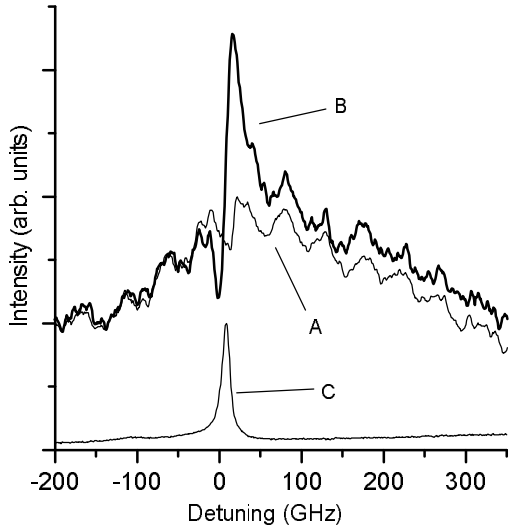}
\caption{\label{fig8}Experiment with narrowband pump; dye-laser spectra;
pulsed discharge afterglow, $P=14$ Torr, $I=4.5$ A, $t_p=100$ $\mu$s,
$\tau_d=50$ $\mu$s, $\lambda=640.2$ nm, $n_0 \approx 10^{13}$ cm$^{-3}$,
$\alpha_0 L \approx 200$, $W=20$ $\mu$J; A, original probe beam; B, the probe
beam after interaction with the medium; C, the pump beam in a ``narrowband''
mode; spectral intensities of probe and pump fields are presented in
different arbitrary units.}
\end{figure}

The dependence of the probe transmission spectrum on the pump detuning
$\Delta_{0}$ was studied under the ``narrowband'' pumping. For these
measurements the dye laser (1), producing the pump beam, was switched to the
narrowband mode (see Sec.~II) with the spectral width [full width at half
maximum (FWHM)] $\gamma_{\rm pump}/2\pi = 10$ GHz. Here, $\Delta_{0}$
corresponds to the detuning between the center of the pump spectrum and the
atomic transition. Typical pump spectrum is shown by the curve C in
Fig.~\ref{fig8}. Comparison of this curve with Fig.~\ref{fig2}(a) explains
the difference between the broadband and narrowband mode.

An earlier version of our experiment was presented in Ref.~\cite{27}.
Amplification spectra, observed in Ref.~\cite{27} under narrowband pumping,
were similar to the first and third types described above. The main
disadvantage of our earlier experiments \cite{25,26,27} was that both the
probe and pump beam were obtained from the same dye laser. Scanning the
frequency of the laser in Ref.~\cite{27} revealed some interesting behavior
of the amplification coefficient value, while the shape of the transmission
spectrum was not affected. In the first amplification regime with the
``resonant peak'' spectrum, amplification is the most effective under
resonant pumping $\Delta_{0} = 0$, while in the third regime with the
dispersionlike spectrum, amplification reaches its maximum, when the pump is
detuned to the blue side from the resonance at $\Delta_{0} = 7$ GHz. The
present experimental setup allows us to scan the pump frequency keeping the
probe spectrum fixed. The results of the present study are in general
agreement with the data reported in Ref.~\cite{27}. Typical result of this
experiment is presented in Fig.~\ref{fig8} demonstrating the dispersionlike
amplification contour.

\subsection{Conical emission}

In the series of measurements with metastable atom density of about $10^{13}$
cm$^{-3}$ (in pulsed discharge afterglow) and great pump beam intensity ($W
\approx 50 - 100$ $\mu$J in a ``broadband'' mode, or about 10 $\mu$J in a
``narrowband'' mode), we observed the amplification of radiation in the
periphery of the pump beam, appeared independently on the probe-beam
presence. This radiation propagates at the angle of $\theta=10$ - 15 mrad
with respect to the pump beam and forms a cone around it. In the literature,
this phenomenon is usually referred to as conical emission. It should be
emphasized that in this domain, all known experiments were carried out in
vapors of alkali and rare-earth elements. Here we present the first conical
emission observed in a noble gas.

\begin{figure}
\includegraphics{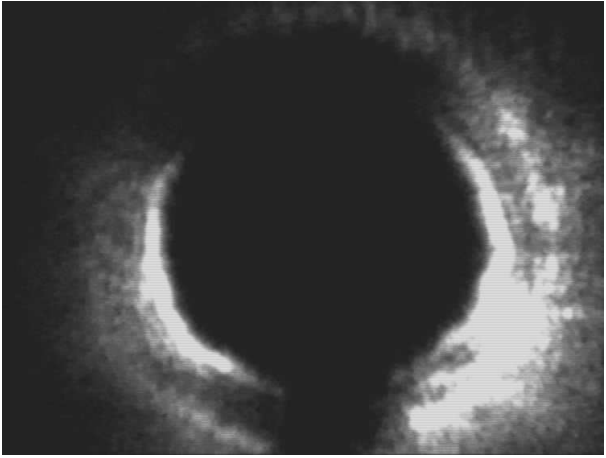}
\caption{\label{fig9}Pump-beam transverse structure at the output of the
discharge tube;  pulsed discharge afterglow, $P=14$ Torr, $I=4.5$ A, $t_p=80$
$\mu$s, $\tau_d=40$ $\mu$s, $\lambda =640.2$ nm, $W=10$ $\mu$J, narrowband
pump, $\Delta_0 =0$; central part of the beam is blocked.}
\end{figure}

The transverse profile of the radiation after its passage through the
discharge tube was recorded by the same CCD camera that was used for the
probe-beam spectrum investigation. For this experiment the camera was
installed at a distance of 1 m behind the discharge tube. The central and the
most intense part of the pump beam was blocked to avoid damage of the camera.
In Fig.~\ref{fig9}, a typical image taken by the camera is shown. The inner
ring in this picture corresponds to the periphery of the pump beam, whereas
the outer one represents the conical emission. In a range of parameters
explored, we did not find any dependence of the cone angle $\theta$ on
$J_{\rm pump}$ or $n_0$. Increase in the time delay $\tau_d$ leads to the
decreasing medium density and cone intensity.

\begin{figure}
\includegraphics{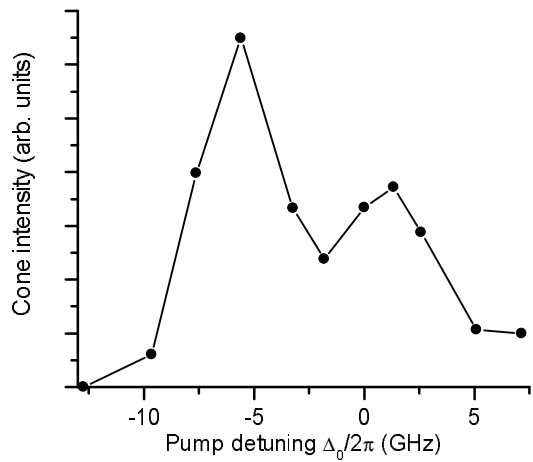}
\caption{\label{fig10}Cone intensity dependence on the pump detuning
$\Delta_0$; pulsed discharge afterglow, $P=14$ Torr, $I=4.5$ A, $t_p =80$
$\mu$s, $\tau_d=20$ $\mu$s, $\lambda =640.2$ nm, $W=10$ $\mu$J, narrowband
pump.}
\end{figure}

We also observed conical emission in the experiments with narrowband pump. In
this regime, the cone intensity depends on the pump detuning $\Delta_0$: it
has two maxima on the wings of the absorption line and disappears at the
large detunings $|\Delta_{0}/2\pi| > 10$ GHz. This dependence is plotted in
Fig.~\ref{fig10}.

\begin{figure}
\includegraphics{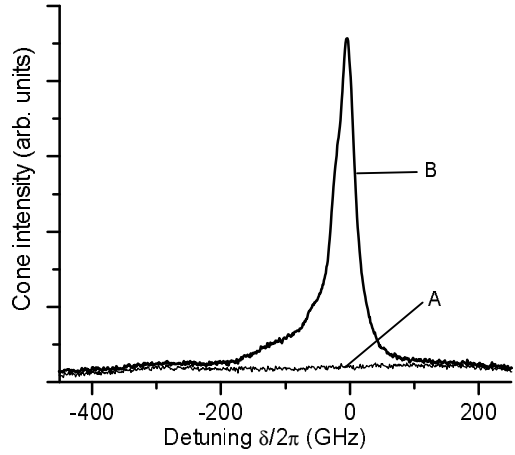}
\caption{\label{fig11}Conical emission spectrum; pulsed discharge afterglow,
$P=9$ Torr, $I=4.0$ A, $t_p=100$ $\mu$s, $\tau_d=16$ $\mu$s, $\lambda =640.2$
nm, $W=53$ $\mu$J, broadband pump;  A, discharge off;  B, discharge on.}
\end{figure}

The conical emission spectra were recorded together with the probe-beam
spectra under the same conditions. We observed spectra with shapes very
similar to the probe-beam amplification spectra (except for the
dispersionlike contour). One example of the conical emission spectrum under
broadband pumping is shown in Fig.~\ref{fig11}. The single peak is situated
near the atomic resonance and is shifted to the red side. This fact is in
agreement with other conical emission experiments, although the width of the
peak and its detuning are different.

\section{Theoretical model and discussion}

\subsection{Equations of the model}

Theoretical model is based on the solution of the semiclassical
Maxwell-Bloch equations in the two-level approximation \cite{28}.
Using the rotating wave approximation, the system of Bloch
equations can be written as

\begin{subequations}\label{1}
\begin{eqnarray}
\frac{\partial p}{\partial t} & = & \Omega D -\gamma_2 p, \\
\frac{\partial D}{\partial t} & = & -\frac{1}{2}\left(\Omega p^* + \Omega^*
p\right)- \gamma_1\left(D-D^{eq}\right),
\end{eqnarray}
\end{subequations}
where $\Omega=2dE/\hbar$ is the complex Rabi frequency of the electromagnetic
field ($d$ is the electric dipole moment of the atomic transition), $p$ and
$D$ are the complex polarization and population difference of atoms, and
$D^{eq}$ is the value of $D$ in the absence of external field (the value
$D=1$ corresponds to an atom in the ground state). The relaxation rates
$\gamma_1$ and $\gamma_2$ determine the homogeneous broadening of a spectral
line. Here $\Omega$ and $p$ are the functions slowly varying in time but
having arbitrary spatial dependence.

The problem of the interaction between two intersected plane linearly
polarized waves was considered. In this case, the amplitude of the field is
given by

\begin{equation}\label{2}
\Omega=\Omega_0(t,z)e^{-ik_{0}z}+\Omega_1(t,z)e^{-i{\bf k}_{1}{\bf r}},
\end{equation}
with $\Omega_0$ and $\Omega_1$ slowly varying in space. The field $\Omega_0$
with a wave vector ${\bf k}_0$ parallel to the $z$ axis corresponds to the
strong pump wave, whereas $\Omega_1$ with ${\bf k}_1$ wave vector is assumed
to be a weak probe field propagating at a small angle of $\varphi$ with
respect to $z$ direction. Nonlinear interaction of the intersected waves
leads to the appearance of spatial polarization harmonics with wave vectors
${\bf k}_0+m\Delta{\bf k}$ ($m=0, \pm 1, \pm 2,..$, $\Delta{\bf k}={\bf
k}_1-{\bf k}_0$) and harmonics of the population difference with $m\Delta{\bf
k}$ wave vectors:

\begin{subequations}\label{3}
\begin{eqnarray}
p=\sum_{m=-\infty}^{\infty}p_m(t,z)e^{-i({\bf k}_{0}{\bf r}+m\Delta{\bf
k}{\bf r})},
\\
D=\sum_{m=-\infty}^{\infty}D_m(t,z)e^{-im\Delta{\bf k}{\bf r}} , \qquad
D_m=D_{-m}^* .
\end{eqnarray}
\end{subequations}
The emission of the $p_0$ and $p_1$ polarizations corresponds to the pump and
probe fields, respectively. The emission of higher harmonics is considered to
be suppressed in a thick medium, due to mismatch in dispersion relation,
which is in agreement with our experimental conditions. Substituting
expansions (\ref{2}) and (\ref{3}) into Eq.~(\ref{1}), and using the
first-order perturbation theory in respect of the small amplitude of the
probe field, one can get a system of the Maxwell-Bloch equations, describing
propagation of the strong pump field,

\begin{subequations}\label{4}
\begin{eqnarray}\
c\frac{\partial \Omega_0}{\partial z}&+&\frac{\partial \Omega_0}{\partial
t}=-\omega_c^2 p_0 , \\
\frac{\partial p_0}{\partial t}&=&\Omega_0
D_0-\gamma_2 p_0,  \\
\frac{\partial D_0}{\partial t}&=&-\frac{1}{2}\left(\Omega_0 p_0^* +
\Omega_0^* p_0\right)- \gamma_1\left(D_0-D^{eq}\right),
\end{eqnarray}
\end{subequations}
and a weak probe,

\begin{subequations}\label{5}
\begin{eqnarray}
c \cos\varphi\frac{\partial \Omega_1}{\partial z}+\frac{\partial
\Omega_1}{\partial t}=-\omega_c^2 p_1 , \\
\frac{\partial p_1}{\partial t}=\Omega_1 D_0+\Omega_0 D_1-\gamma_2 p_1, \\
\frac{\partial D_1}{\partial t}=-\frac{1}{2}\left(\Omega_1 p_0^* +
\Omega_0^* p_1+\Omega_0 p_{-1}^*\right)- \gamma_1 D_1, \\
\frac{\partial p_{-1}^*}{\partial t}=\Omega_0^* D_1-\gamma_2 p_{-1}^*,
\end{eqnarray}
\end{subequations}
where

\begin{equation}\label{6}
\omega_c=\sqrt{\frac{2\pi d^2 \omega_0 n_0}{\hbar}}
\end{equation}
is the cooperative frequency of the medium, which plays a role of the
coupling coefficient between field and matter, and $n_0$ is the density of
atoms in the ground state.

The character of resonant coherent interaction between laser field and an
ensemble of two-level atoms is determined by the atomic density and
particular characteristics of laser pulse, such as the area and energy.
Temporal features of propagating pulse strongly depend on the condition
whether the field reemitted by the medium is comparable to the field that has
been externally applied. In the limiting case, where reemission field is
negligible in comparison to the strong externally applied one, the model of
single driven atom is usually used. Atomic response is calculated on the
basis of Bloch equations, whereas Maxwell equations are not taken into
account. Thus, the dynamics of the system is fully determined by the strong
external field. In this context, phenomena such as Rabi sideband generation
due to stationary probe-pump Mollow-Boyd effect \cite{10,11,12} or to
transient Rabi flopping \cite{28} can be mentioned.

In a system, where the field of medium reaction plays an essential role,
Bloch and Maxwell equations are considered self-consistently. In the case of
a single (pump) field propagation, under the simplest conditions of an
amplitude modulated signal, and in the limit of coherent interaction
$\gamma_{1,2}=0$, Eqs. (\ref{4}) can be reduced to the nonlinear evolution
sine-Gordon equation, which describes the coherent breakup of a laser pulse
into solitons ($2\pi$ pulses or breathers) \cite{29} and a nonsolitonic
solution in the form of the so-called optical ringing \cite{30}. $2\pi$
solitons of self-induced transparency represent the pulses, where the
coherently absorbed and reemitted fields are balanced. Nevertheless, in this
case, the field applied is still strong enough to invert a part of a medium
(area of the external field is greater than $\pi$).

In accordance with the area theorem \cite{28}, if the area of the input field
takes the value less than $\pi$, the nonstationary dynamics of the pulse
breakup into $2\pi$ solitons cannot be realized. On the other hand, in an
optically dense coherent medium, another type of field oscillations, optical
ringing, appears. This oscillating response has a superradiant character
\cite{4} and appears in the system of strongly coupled field and matter,
where the high density of resonant atoms provides the atomic system to be
able to coherently absorb and reradiate whole energy of the weak external
electromagnetic field. The oscillations display the process of fast
excitation interchanges between field and a two-level atomic ensemble in an
extended resonant medium and the formation of a $0\pi$ pulse. Such
cooperative interactions are the transient phenomena and can be observed when
the frequency of the photon interchanges between field and matter greatly
exceeds relaxation rates of a medium.

The collective oscillations can be treated as beating between two normal
modes of coupled linear oscillators, which correspond to splitting of the
polariton (coupled matter photons) dispersion curve into two branches. Thus,
contrary to nonlinear strong-field effects, the coherent dynamics of strongly
coupled field-matter system originates from linear (weak-field) dispersion,
where field and polarization components are represented by equal
contributions, and reemission field plays an essential role in the
light-matter interaction. Vacuum Rabi oscillations (both single- and
many-atom oscillations) \cite{2,3} are cavity analogs of such processes. In
the simplest case of a single-mode cavity, the frequency of field and
polarization oscillations is equal to the cooperative frequency of the
medium, $\omega_c$, when the condition of strong-coupling regime $\omega_c
\gg \gamma_2$ is fulfilled. Under free-space interactions, the frequency of
field-matter photon interchanges is also determined by the coupling
coefficient (6). To observe coherent beating of two (red and blue) polariton
branches, they should be excited by the laser field simultaneously. In our
experiment, this condition is fulfilled due to resonant broadband (much
broader than the Doppler linewidth) spectrum of the radiation used.

\subsection{Numerical results}

We studied numerically interaction of polychromatic broadband
quasistochastic laser pulses in a dense medium without population
inversion. The parameters of numerical simulations were similar to
the experimental conditions. Input pump field was chosen to have
the following form:

\begin{eqnarray}\label{7}
\Omega_0(t,0)=C(t)\sum_{k=-N}^{k=N}\Omega_{0k}e^{i\left(\Delta_0 t
+\omega_k t +\alpha_k\right)} , \qquad \nonumber\\
\Omega_{0k}=\Omega_{00}e^{-4\ln{2}
{(\omega_k /\gamma_{\rm pump})}^2} , \qquad \nonumber\\
C(t)=\frac{2}{\pi}e^{-(t-t_0)/a}\left[\frac{\pi}{2}+\arctan
\left({\frac{t-t_0}{b}}\right)\right],
\end{eqnarray}
where $\omega_k=k\Delta\omega$ ($k=0,\pm 1,..., \pm N$) are the modes of the
input spectrum, $\Delta\omega$ is the intermode distance, $\Delta_0$ is the
detuning between the central frequency of the field and that of the atomic
resonance, and $\alpha_k$ are the mode phases. The phases are random numbers,
which leads to the quasistochastic temporal dependence of the electric field.
The duration of this signal does not depend on the width of the spectrum
$\gamma_{\rm pump}$, but is determined by the duration of the envelope,
$C(t)$.

Modeling one of the experimental realizations, we used the following
parameters of the input signal: $\Delta\omega/2\pi=0.37$ GHz, $\gamma_{\rm
pump}/2\pi=20.0$ GHz (thus, the number of modes was about 100), the duration
of the envelope, $C(t)$, was equal to 3.5 ns ($a=2.4$ ns, $b=0.3$ ns). The
amplitude of the probe field has the form $\Omega_1(t,
0)=\Omega_0(t-\tau_0,0)/g$, where $g\gg 1$, and a short (about 10 ps) time
delay $\tau_0$ between the fields was taken into account. The atomic
transition with $\gamma_1/2\pi=8.4$ MHz and $\gamma_2/2\pi=5.4$ MHz was
considered. The cooperative frequency of the medium was equal to $\omega_c
/2\pi=2.6$ GHz and the length of the medium $L=15$ cm. Since correlation time
of the field was much smaller than the pulse duration, and great number of
modes were considered, main spectral features of the output field were not
sensitive to the certain realization of random phases and a small time delay
$\tau_0$ existing in the experiment.

\begin{figure}
\includegraphics{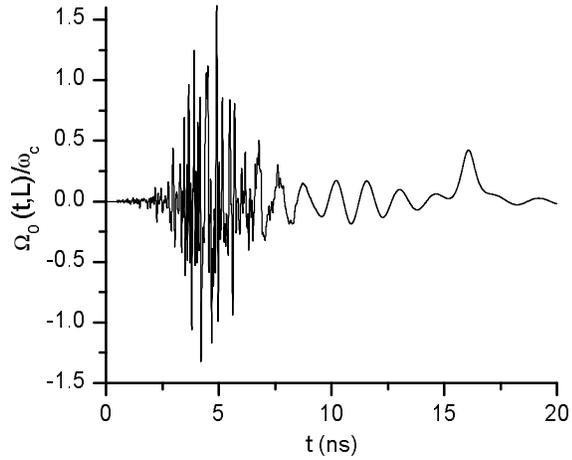}
\caption{\label{fig12}Temporal behavior of a polychromatic pulse after
propagation in an extended resonant medium; numerical modeling, $\omega_c
/2\pi=2.6$ GHz, $L=15$ cm.}
\end{figure}

A typical example of the temporal behavior of the single (pump) field at the
output of the extended medium is presented in Fig.~\ref{fig12}. The figure
shows that during the propagation of the multimode radiation (\ref{7})
through a resonant extended medium, smooth slow solitons separate from the
quasistochastic part, that forms the initial stage of the signal. The optical
ringing accompanies the creation of a soliton and forms the tail of the light
pulse. If the area of the input pulse takes the value less than $\pi$, $2\pi$
solitons are not created and the optical ringing signal can be the only and
dominant part at the pulse tail. Thus, Fig.~\ref{fig12} shows the
transformation of the quasistochastic signal with correlation time determined
by the width of the input spectrum, $\gamma_{\rm pump}$ (\ref{7}), into the
coherent response of the dense resonant medium, determined mainly by the
field-matter coupling coefficient $\omega_c$ (\ref{6}) and by the length of
the medium, $L$.

\begin{figure}
\includegraphics{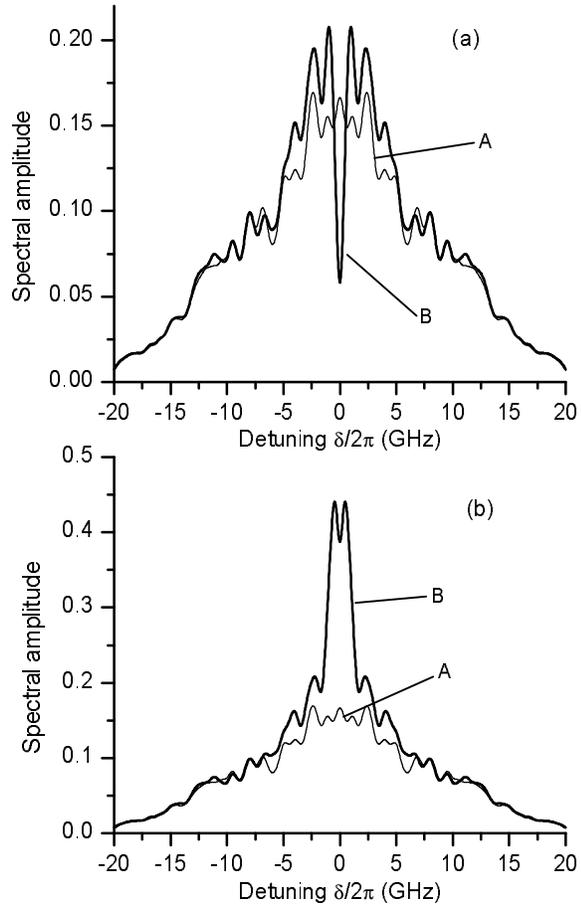}
\caption{\label{fig13}Calculated spectra of the probe field at the input
(curves A) and output (curves B) of a medium; (a) doublet amplification, (b)
amplification with resonant maximum; $\omega_c /2\pi=2.6$ GHz, $L=15$ cm.}
\end{figure}

The numerical study was based on the joint solution of Eqs.~(\ref{4}) and
(\ref{5}). In Fig.~\ref{fig13}, spectra of the probe field at the input and
output of the medium are displayed. The spectra presented were obtained by
convoluting the calculated spectra with a smooth Gaussian function, modeling
the transmission of the signal through a device with finite resolution, so
single modes of the polychromatic signal are not resolved. Depending on the
pump intensity, two types of the probe amplification were obtained.

If the pump field is strong enough to produce nonstationary population
inversion in a two-level extended medium, the probe is significantly
amplified at the resonance frequency [Fig.~\ref{fig13}(b)]. The most
essential growth of the probe field is obtained in the space and time areas,
where a smooth $2\pi$ soliton has formed in the pump beam, which leads to the
slow rotation of the Bloch vector and smooth variation of the population
difference between $1$ and $-1$ values, which corresponds to transitions of
the atoms from the ground state to the excited one and back through the
coherent superposition states.

In the case of relatively small pumping intensity, the pump pulse can
propagate without producing a soliton. This condition leads to oscillations
of the Bloch vector near the equilibrium point $D=1$. Here, the amplification
spectrum obtained corresponds to the spectral doublet presented in
Fig.~\ref{fig13}(a). Moreover, a slow soliton separated from the main fast
part of the signal can be absorbed in the medium during the propagation. As a
result, even in this case, amplification is formed by the fast part of the
signal (quasistochastic and ringing), which has the area equal to zero,
which, again, corresponds to the doublet in the amplification spectrum. As
was shown in Refs.~\cite{25,26} (see also Ref.~\cite{38}), under some
conditions, the interaction of $0\pi$ pulses in a dense medium results in
amplification of the coherent optical ringing. Besides that, two maxima in
the amplification spectrum can be parametrically coupled due to the
modulation of the population difference by the pump field.

The doublet and single-maximum characters of the probe
amplification under the weak and strong pumping, respectively,
qualitatively correspond to our experimental observations
presented in Figs.~\ref{fig3} and \ref{fig4}. The origin of such
spectral behavior of the probe field can be traced to the dynamics
of the pump. If the pump pulse has the zero area, according to the
area theorem, its output spectrum has a form of a doublet, whereas
for stronger pulses with the areas greater than $\pi$, the
resonant spectral component (which is equal to the area) takes a
nonzero value.

The single-maximum amplification that is obtained from the numerical
simulations can be greater than the one observed in the experiments. The
discrepancy can be explained by the fact, that this type of amplification
appears very close to the resonance frequency, since the spectral width of a
smooth soliton is much narrower than that of the broadband polychromatic part
(cf. Fig.~\ref{fig12}). So, the amplification maximum can be sufficiently
reduced by the Doppler broadening of the spectral line, which was not taken
into account in the theoretical model presented. On the other hand, the width
of the doublet amplification, which appears due to the fast oscillations, can
take a value greater than the Doppler linewidth.

\subsection{Strong probe beam and conical emission}

Our numerical model describes the interaction of two intersected
plane waves. Therefore it does not account for some transverse
effects, such as self-focusing of a beam and conical emission.
Besides that, since the probe field was treated only in the first
order, the model does not describe nonlinear effects in the probe
beam.

In the experiment, increase in the probe beam power leads to some asymmetry
in the transmission spectrum. In this case, even without the pump beam, the
probe beam intensity is sufficient to induce nonlinear effects, which was
stressed is Sec. III A. As the pulse duration is shorter than the homogeneous
relaxation times, self-induced transparency can dramatically modify the
absorption spectrum. It has been shown numerically \cite{31,32} and proved
experimentally \cite{23} that presence of a nonzero detuning $\delta_0$ such
that $\gamma_D < \delta_0 < \gamma_{\rm probe}/2$ results in the appearance
of a dispersionlike feature centered at the transition frequency. The
spectral components with detunings of the same sign as $\delta_0$ are
amplified, and those with opposite sign are attenuated. We suppose that a
transmission spectrum plotted in Fig.~\ref{fig5} (curve B) is related to that
effect. In our experiment with pulsed discharge, the evolution of the
dispersionlike feature under the increase in the metastable atom density,
which was described in Sec. III A (measurements were carried out in the
presence of the pump beam, its effect is discussed below) is in agreement
with numerical results of Ref.~\cite{32}. More complicated spectral
structure, obtained in Refs.~\cite{23,32} under great optical densities, was
unresolved by our apparatus.

Applying a strong pump causes the probe-beam amplification and increases the
nonlinear features in its spectrum. It leads to the increase in the probe
intensity integrated over the whole spectrum and increases the asymmetry of
the transmission spectrum. This effect is clearly seen in Fig.~\ref{fig5}.

With a single probe beam, we observed dispersionlike spectra of both ``left''
and ``right'' orientations, in agreement with Refs.~\cite{31,32}. In the
presence of the pump beam, a structure with amplification at the blue wing
certainly prevails over the opposite one. This fact may be understood, taking
into account the effect of resonant self-focusing of the pump beam, which is
well pronounced under these experimental conditions. The problem of resonant
and near resonant self-focusing and its influence on the propagation of SIT
pulses was discussed mostly for relatively narrowband pulses ($\gamma_{\rm
pulse} \approx \gamma_D$). It has been shown numerically \cite{33} and
experimentally \cite{34} that during propagation in an extended medium,
resonant and slightly blue-detuned SIT pulses become self-focused, while
red-detuned pulses experience self-defocusing. This effect strongly modifies
transverse profile of the pump beam as well as the profile of the probe, when
its intensity becomes sufficiently high. Self-focusing leads to an increase
in the axial intensity of the blue components of the probe that increases the
effective detuning $\delta_0$ of the central, the most intense part of the
beam, and results in more pronounced asymmetry of the transmission spectrum.
The interplay between parametric amplification, self- and pump-induced
focusing of the probe beam, and propagation effects creates very complex
picture, which may be quantitatively described only by numerical modeling.

Conical emission observed in the present experiment appeared under the
conditions that are significantly different from other CE experiments. We
obtain CE under the metastable atom density of about $10^{13}$ cm$^{-3}$ and
the pump detunings $\Delta_0$ (see Fig.~\ref{fig10}) comparable to the
Doppler width of the transition, whereas ``normal'' CE is observed under much
greater detunings and mostly under higher medium densities. Besides that, CE
observed in the present work is insensitive to the sign of the pump detuning
$\Delta_0$, whereas ``normal'' CE appears under blue-detuned pump only. The
model of Cherenkov-type emission from steady-state self-trapped filaments
\cite{18}, as well as the steady-state four-wave mixing model \cite{19}, seem
to be unsuitable under conditions of our experiment. As a nonstationary
coherent phenomenon, CE was considered in Refs.~\cite{35,36,37}. In these
works dealing with transform limited pulses, CE was obtained under the
breaking up of a laser pulse, which at least qualitatively may also
correspond to our results. The shape of CE spectrum presented here is similar
to the amplified probe-beam spectrum observed under the same conditions. We
suppose that the mechanisms of these two phenomena are closely related.

\section{Conclusion}

The detailed investigation of changes in the spectrum of a polychromatic
probe field, which arise during its propagation in a dense resonant medium,
is presented. The spectral width of the radiation considered greatly exceeds
both the homogeneous and inhomogeneous widths of the spectral line. In the
simplest case of a single-beam propagation, the dispersionlike asymmetry of
the transmission spectrum was observed, which was treated as a feature of
near resonant self-induced transparency in the so-called sharp line limit.
When two laser beams are intersected in a medium, we observed dramatic
changes in the probe-beam spectrum: pump-induced amplification and
attenuation. The shape of the transmission spectrum depends on the pump and
probe intensities, pump detuning, and atomic density. Under relatively weak
pumping and high medium density, when the condition of strong coupling
between field and resonant matter is fulfilled, the probe amplification
spectrum has a form of spectral doublet. Stronger pumping leads to the
appearance of a single peak of the probe beam amplification at the transition
frequency, manifesting that strong-field effects prevail over effects in a
strongly coupled field-matter system. The greater probe intensity results in
an asymmetrical transmission spectrum with amplification at the blue wing of
the absorption line and attenuation at the red one. Under high medium
density, a broad band of amplification appears. Since, in the situation
considered, the field of medium reaction plays an important role and forms a
significant part of the coherent collective response of a dense medium, the
amplification spectra observed differ from the well-known Mollow-Boyd
spectra, which are determined by characteristics of the strong external
field. The theoretical model presented is based on the numerical solution of
the Maxwell-Bloch equations for two plane waves propagating in a dense
extended medium. It is shown that the characteristic features of the probe
amplification could be determined by the coherent nonstationary dynamics of
the pump field, particularly by the formation of solitons and optical
ringing. In some cases, the pump-beam transverse profile also should be taken
into account. Under the certain conditions, the periphery area of the pump
beam becomes amplified and produces a conical emission.  Some parameters of
CE obtained in the present work differ from usually observed CE. It should be
pointed out that amplified probe-beam spectra reproduce the most
characteristic features of generation spectrum of a broadband laser with an
intracavity absorbing cell, when the self-frequency-locking effect, i.e.,
spectrum condensation, takes place \cite{8,9}.

\begin{acknowledgments}
The work was partially supported by the INTAS, Project No. 99-1366.
\end{acknowledgments}

\begin{thebibliography}{38}
\expandafter\ifx\csname natexlab\endcsname\relax\def\natexlab#1{#1}\fi
\expandafter\ifx\csname bibnamefont\endcsname\relax
  \def\bibnamefont#1{#1}\fi
\expandafter\ifx\csname bibfnamefont\endcsname\relax
  \def\bibfnamefont#1{#1}\fi
\expandafter\ifx\csname citenamefont\endcsname\relax
  \def\citenamefont#1{#1}\fi
\expandafter\ifx\csname url\endcsname\relax
  \def\url#1{\texttt{#1}}\fi
\expandafter\ifx\csname urlprefix\endcsname\relax\def\urlprefix{URL }\fi
\providecommand{\bibinfo}[2]{#2} \providecommand{\eprint}[2][]{\url{#2}}

\bibitem[{\citenamefont{Zheleznyakov et~al.}(1989)\citenamefont{Zheleznyakov,
  Kocharovsky, and Kocharovsky}}]{1}
\bibinfo{author}{\bibfnamefont{V.~V.} \bibnamefont{Zheleznyakov}},
  \bibinfo{author}{\bibfnamefont{V.~V.} \bibnamefont{Kocharovsky}},
  \bibnamefont{and} \bibinfo{author}{\bibfnamefont{Vl.~V.}
  \bibnamefont{Kocharovsky}}, \bibinfo{journal}{Usp.\ Fiz.\ Nauk}
  \textbf{\bibinfo{volume}{159}}, \bibinfo{pages}{193} (\bibinfo{year}{1989})
  [\bibinfo{journal}{Sov.\ Phys.\ Usp.} \textbf{\bibinfo{volume}{32}},
  \bibinfo{pages}{835} (\bibinfo{year}{1989})].

\bibitem[{\citenamefont{Kaluzny et~al.}(1983)\citenamefont{Kaluzny,
  Goy, Gross, Raimond, and Haroche}}]{2}
\bibinfo{author}{\bibfnamefont{Y.} \bibnamefont{Kaluzny}},
  \bibinfo{author}{\bibfnamefont{P.} \bibnamefont{Goy}},
  \bibinfo{author}{\bibfnamefont{M.} \bibnamefont{Gross}},
  \bibinfo{author}{\bibfnamefont{J.~P.} \bibnamefont{Raimond}},
  \bibnamefont{and} \bibinfo{author}{\bibfnamefont{S.}
  \bibnamefont{Haroche}}, \bibinfo{journal}{Phys.\ Rev.\ Lett.}
  \textbf{\bibinfo{volume}{51}}, \bibinfo{pages}{1175}
  (\bibinfo{year}{1983}).

\bibitem[{\citenamefont{Zhu et~al.}(1990)\citenamefont{Zhu,
  Gauthier, Morin, Wu, Carmichael, and Mossberg}}]{3}
\bibinfo{author}{\bibfnamefont{Y.} \bibnamefont{Zhu}},
  \bibinfo{author}{\bibfnamefont{D.~J.} \bibnamefont{Gauthier}},
  \bibinfo{author}{\bibfnamefont{S.~E.} \bibnamefont{Morin}},
  \bibinfo{author}{\bibfnamefont{Q.} \bibnamefont{Wu}},
  \bibinfo{author}{\bibfnamefont{H.~J.} \bibnamefont{Carmichael}},
  \bibnamefont{and} \bibinfo{author}{\bibfnamefont{T.~W.}
  \bibnamefont{Mossberg}}, \bibinfo{journal}{Phys.\ Rev.\ Lett.}
  \textbf{\bibinfo{volume}{64}}, \bibinfo{pages}{2499}
  (\bibinfo{year}{1990}).

\bibitem[{\citenamefont{Prasad and Glauber}(2000)}]{4}
\bibinfo{author}{\bibfnamefont{S.}~\bibnamefont{Prasad}} \bibnamefont{and}
  \bibinfo{author}{\bibfnamefont{R.~J.} \bibnamefont{Glauber}},
  \bibinfo{journal}{Phys.\ Rev.\ A} \textbf{\bibinfo{volume}{61}},
  \bibinfo{pages}{063814} (\bibinfo{year}{2000}).

\bibitem[{\citenamefont{Huynh et~al.}(2002)\citenamefont{Huynh, Tignon,
  Roussignol, Delalande, Andre, Romestain, and Dang}}]{5}
\bibinfo{author}{\bibfnamefont{A.}~\bibnamefont{Huynh}},
  \bibinfo{author}{\bibfnamefont{J.}~\bibnamefont{Tignon}},
  \bibinfo{author}{\bibfnamefont{P.}~\bibnamefont{Roussignol}},
  \bibinfo{author}{\bibfnamefont{C.}~\bibnamefont{Delalande}},
  \bibinfo{author}{\bibfnamefont{R.}~\bibnamefont{Andre}},
  \bibinfo{author}{\bibfnamefont{R.}~\bibnamefont{Romestain}},
  \bibnamefont{and} \bibinfo{author}{\bibfnamefont{D.~L.~S.}
  \bibnamefont{Dang}}, \bibinfo{journal}{Phys.\ Rev.\ B}
  \textbf{\bibinfo{volume}{66}}, \bibinfo{pages}{113301}
  (\bibinfo{year}{2002}).

\bibitem[{\citenamefont{Messin et~al.}(2001)\citenamefont{Messin, Karr, Baas,
  Khitrova, Houdre, Stanley, Oesterle, and Giacobino}}]{6}
\bibinfo{author}{\bibfnamefont{G.}~\bibnamefont{Messin}},
  \bibinfo{author}{\bibfnamefont{J.~P.} \bibnamefont{Karr}},
  \bibinfo{author}{\bibfnamefont{A.}~\bibnamefont{Baas}},
  \bibinfo{author}{\bibfnamefont{G.}~\bibnamefont{Khitrova}},
  \bibinfo{author}{\bibfnamefont{R.}~\bibnamefont{Houdre}},
  \bibinfo{author}{\bibfnamefont{R.~P.} \bibnamefont{Stanley}},
  \bibinfo{author}{\bibfnamefont{U.}~\bibnamefont{Oesterle}}, \bibnamefont{and}
  \bibinfo{author}{\bibfnamefont{E.}~\bibnamefont{Giacobino}},
  \bibinfo{journal}{Phys.\ Rev.\ Lett.} \textbf{\bibinfo{volume}{87}},
  \bibinfo{pages}{127403} (\bibinfo{year}{2001}).

\bibitem[{\citenamefont{Savvidis et~al.}(2000)\citenamefont{Savvidis, Baumberg,
  Stevenson, Skolnick, Whittaker, and Roberts}}]{7}
\bibinfo{author}{\bibfnamefont{P.~G.} \bibnamefont{Savvidis}},
  \bibinfo{author}{\bibfnamefont{J.~J.} \bibnamefont{Baumberg}},
  \bibinfo{author}{\bibfnamefont{R.~M.} \bibnamefont{Stevenson}},
  \bibinfo{author}{\bibfnamefont{M.~S.} \bibnamefont{Skolnick}},
  \bibinfo{author}{\bibfnamefont{D.~M.} \bibnamefont{Whittaker}},
  \bibnamefont{and} \bibinfo{author}{\bibfnamefont{J.~S.}
  \bibnamefont{Roberts}}, \bibinfo{journal}{Phys.\ Rev.\ Lett.}
  \textbf{\bibinfo{volume}{84}}, \bibinfo{pages}{1547} (\bibinfo{year}{2000}).

\bibitem[{\citenamefont{Meyer}(1979)}]{8}
\bibinfo{author}{\bibfnamefont{Y.~H.} \bibnamefont{Meyer}},
  \bibinfo{journal}{Opt.\ Commun.} \textbf{\bibinfo{volume}{30}},
  \bibinfo{pages}{75} (\bibinfo{year}{1979}).

\bibitem[{\citenamefont{Vasil'ev et~al.}(1994)\citenamefont{Vasil'ev, Egorov,
  Fedorov, and Chekhonin}}]{9}
\bibinfo{author}{\bibfnamefont{V.~V.} \bibnamefont{Vasil'ev}},
  \bibinfo{author}{\bibfnamefont{V.~S.} \bibnamefont{Egorov}},
  \bibinfo{author}{\bibfnamefont{A.~N.} \bibnamefont{Fedorov}},
  \bibnamefont{and} \bibinfo{author}{\bibfnamefont{I.~A.}
  \bibnamefont{Chekhonin}}, \bibinfo{journal}{Opt.\ Spectrosk.}
  \textbf{\bibinfo{volume}{76}}, \bibinfo{pages}{146} (\bibinfo{year}{1994})
  [\bibinfo{journal}{Opt.\ Spectrosc.}
  \textbf{\bibinfo{volume}{76}}, \bibinfo{pages}{134} (\bibinfo{year}{1994})].

\bibitem[{\citenamefont{Rautian and Sobel'man}(1961)}]{10}
\bibinfo{author}{\bibfnamefont{S.~G.} \bibnamefont{Rautian}} \bibnamefont{and}
  \bibinfo{author}{\bibfnamefont{I.~I.} \bibnamefont{Sobel'man}},
  \bibinfo{journal}{Zh.\ Eksp.\ Teor.\ Fiz.} \textbf{\bibinfo{volume}{41}},
  \bibinfo{pages}{456} (\bibinfo{year}{1961}) [\bibinfo{journal}{Sov.\ Phys.\
  JETP} \textbf{\bibinfo{volume}{14}},
  \bibinfo{pages}{328} (\bibinfo{year}{1962})].

\bibitem[{\citenamefont{Mollow}(1972)}]{11}
\bibinfo{author}{\bibfnamefont{B.~R.} \bibnamefont{Mollow}},
  \bibinfo{journal}{Phys.\ Rev.\ A} \textbf{\bibinfo{volume}{5}},
  \bibinfo{pages}{2217} (\bibinfo{year}{1972}).

\bibitem[{\citenamefont{Boyd et~al.}(1981)\citenamefont{Boyd, Raymer, Narum,
  and Harter}}]{12}
\bibinfo{author}{\bibfnamefont{R.~W.} \bibnamefont{Boyd}},
  \bibinfo{author}{\bibfnamefont{M.~G.} \bibnamefont{Raymer}},
  \bibinfo{author}{\bibfnamefont{P.}~\bibnamefont{Narum}}, \bibnamefont{and}
  \bibinfo{author}{\bibfnamefont{D.~J.} \bibnamefont{Harter}},
  \bibinfo{journal}{Phys.\ Rev.\ A} \textbf{\bibinfo{volume}{24}},
  \bibinfo{pages}{411} (\bibinfo{year}{1981}).

\bibitem[{\citenamefont{Harter et~al.}(1981)\citenamefont{Harter, Narum,
  Raymer, and Boyd}}]{13}
\bibinfo{author}{\bibfnamefont{D.~J.} \bibnamefont{Harter}},
  \bibinfo{author}{\bibfnamefont{P.}~\bibnamefont{Narum}},
  \bibinfo{author}{\bibfnamefont{M.~G.} \bibnamefont{Raymer}},
  \bibnamefont{and} \bibinfo{author}{\bibfnamefont{R.~W.} \bibnamefont{Boyd}},
  \bibinfo{journal}{Phys.\ Rev.\ Lett.} \textbf{\bibinfo{volume}{46}},
  \bibinfo{pages}{1192} (\bibinfo{year}{1981}).

\bibitem[{\citenamefont{Chalupczak et~al.}(1994)\citenamefont{Chalupczak,
  Gawlik, and Zachorowski}}]{14}
\bibinfo{author}{\bibfnamefont{W.}~\bibnamefont{Chalupczak}},
  \bibinfo{author}{\bibfnamefont{W.}~\bibnamefont{Gawlik}}, \bibnamefont{and}
  \bibinfo{author}{\bibfnamefont{J.}~\bibnamefont{Zachorowski}},
  \bibinfo{journal}{Phys.\ Rev.\ A} \textbf{\bibinfo{volume}{49}},
  \bibinfo{pages}{4895} (\bibinfo{year}{1994}).

\bibitem[{\citenamefont{Wu et~al.}(1977)\citenamefont{Wu, Ezekiel, Ducloy, and
  Mollow}}]{15}
\bibinfo{author}{\bibfnamefont{F.~Y.} \bibnamefont{Wu}},
  \bibinfo{author}{\bibfnamefont{S.}~\bibnamefont{Ezekiel}},
  \bibinfo{author}{\bibfnamefont{M.}~\bibnamefont{Ducloy}}, \bibnamefont{and}
  \bibinfo{author}{\bibfnamefont{B.~R.} \bibnamefont{Mollow}},
  \bibinfo{journal}{Phys.\ Rev.\ Lett.} \textbf{\bibinfo{volume}{38}},
  \bibinfo{pages}{1077} (\bibinfo{year}{1977}).

\bibitem[{\citenamefont{Weisman et~al.}(2000)\citenamefont{Weisman,
  Wilson-Gordon, and Friedmann}}]{16}
\bibinfo{author}{\bibfnamefont{P.}~\bibnamefont{Weisman}},
  \bibinfo{author}{\bibfnamefont{A.~D.} \bibnamefont{Wilson-Gordon}},
  \bibnamefont{and}
  \bibinfo{author}{\bibfnamefont{H.}~\bibnamefont{Friedmann}},
  \bibinfo{journal}{Phys.\ Rev.\ A} \textbf{\bibinfo{volume}{61}},
  \bibinfo{pages}{053816} (\bibinfo{year}{2000}).

\bibitem[{\citenamefont{Grischkowsky}(1970)}]{17}
\bibinfo{author}{\bibfnamefont{D.}~\bibnamefont{Grischkowsky}},
  \bibinfo{journal}{Phys.\ Rev.\ Lett.} \textbf{\bibinfo{volume}{24}},
  \bibinfo{pages}{866} (\bibinfo{year}{1970}).

\bibitem[{\citenamefont{Paul et~al.}(2002)\citenamefont{Paul, Cooper,
  Gallagher, and Raymer}}]{18}
\bibinfo{author}{\bibfnamefont{B.~D.} \bibnamefont{Paul}},
  \bibinfo{author}{\bibfnamefont{J.}~\bibnamefont{Cooper}},
  \bibinfo{author}{\bibfnamefont{A.}~\bibnamefont{Gallagher}},
  \bibnamefont{and} \bibinfo{author}{\bibfnamefont{M.~G.}
  \bibnamefont{Raymer}}, \bibinfo{journal}{Phys.\ Rev.\ A}
  \textbf{\bibinfo{volume}{66}}, \bibinfo{pages}{063816}
  (\bibinfo{year}{2002}).

\bibitem[{\citenamefont{Valley et~al.}(1990)\citenamefont{Valley, Khitrova,
  Gibbs, Grantham, and Jiainin}}]{19}
\bibinfo{author}{\bibfnamefont{F.}~\bibnamefont{Valley}},
  \bibinfo{author}{\bibfnamefont{G.}~\bibnamefont{Khitrova}},
  \bibinfo{author}{\bibfnamefont{H.~M.} \bibnamefont{Gibbs}},
  \bibinfo{author}{\bibfnamefont{J.~W.} \bibnamefont{Grantham}},
  \bibnamefont{and} \bibinfo{author}{\bibfnamefont{X.}~\bibnamefont{Jiainin}},
  \bibinfo{journal}{Phys.\ Rev.\ Lett.} \textbf{\bibinfo{volume}{64}},
  \bibinfo{pages}{2362} (\bibinfo{year}{1990}).

\bibitem[{\citenamefont{Meyer}(1980)}]{20}
\bibinfo{author}{\bibfnamefont{Y.~H.} \bibnamefont{Meyer}},
  \bibinfo{journal}{Opt.\ Commun.} \textbf{\bibinfo{volume}{34}},
  \bibinfo{pages}{439} (\bibinfo{year}{1980}).

\bibitem[{\citenamefont{Hart et~al.}(1994)\citenamefont{Hart, You, Gallagher,
  and Cooper}}]{21}
\bibinfo{author}{\bibfnamefont{R.~C.} \bibnamefont{Hart}},
  \bibinfo{author}{\bibfnamefont{L.}~\bibnamefont{You}},
  \bibinfo{author}{\bibfnamefont{A.}~\bibnamefont{Gallagher}},
  \bibnamefont{and} \bibinfo{author}{\bibfnamefont{J.}~\bibnamefont{Cooper}},
  \bibinfo{journal}{Opt.\ Commun.} \textbf{\bibinfo{volume}{111}},
  \bibinfo{pages}{331} (\bibinfo{year}{1994}).

\bibitem[{\citenamefont{Ben-Aryeh}(1997)}]{22}
\bibinfo{author}{\bibfnamefont{Y.}~\bibnamefont{Ben-Aryeh}},
  \bibinfo{journal}{Phys.\ Rev.\ A} \textbf{\bibinfo{volume}{56}},
  \bibinfo{pages}{854} (\bibinfo{year}{1997}).

\bibitem[{\citenamefont{Ranka et~al.}(1998)\citenamefont{Ranka, Schirmer, and
  Gaeta}}]{23}
\bibinfo{author}{\bibfnamefont{J.~K.} \bibnamefont{Ranka}},
  \bibinfo{author}{\bibfnamefont{R.~W.} \bibnamefont{Schirmer}},
  \bibnamefont{and} \bibinfo{author}{\bibfnamefont{A.~L.} \bibnamefont{Gaeta}},
  \bibinfo{journal}{Phys.\ Rev.\ A} \textbf{\bibinfo{volume}{57}},
  \bibinfo{pages}{R36} (\bibinfo{year}{1998}).

\bibitem[{\citenamefont{Schupper et~al.}(1999)\citenamefont{Schupper,
  Friedmann, Matusovsky, Rosenbluh, and Wilson-Gordon}}]{24}
\bibinfo{author}{\bibfnamefont{N.}~\bibnamefont{Schupper}},
  \bibinfo{author}{\bibfnamefont{H.}~\bibnamefont{Friedmann}},
  \bibinfo{author}{\bibfnamefont{M.}~\bibnamefont{Matusovsky}},
  \bibinfo{author}{\bibfnamefont{M.}~\bibnamefont{Rosenbluh}},
  \bibnamefont{and} \bibinfo{author}{\bibfnamefont{A.~D.}
  \bibnamefont{Wilson-Gordon}}, \bibinfo{journal}{J.\ Opt.\ Soc.\ Am.\ B}
  \textbf{\bibinfo{volume}{16}}, \bibinfo{pages}{1127} (\bibinfo{year}{1999}).

\bibitem[{\citenamefont{Bagaev
  et~al.}(2002{\natexlab{a}})\citenamefont{Bagaev, Egorov, Mekhov, Moroshkin,
  and Chekhonin}}]{25}
\bibinfo{author}{\bibfnamefont{S.~N.} \bibnamefont{Bagaev}},
  \bibinfo{author}{\bibfnamefont{V.~S.} \bibnamefont{Egorov}},
  \bibinfo{author}{\bibfnamefont{I.~B.} \bibnamefont{Mekhov}},
  \bibinfo{author}{\bibfnamefont{P.~V.} \bibnamefont{Moroshkin}},
  \bibnamefont{and} \bibinfo{author}{\bibfnamefont{I.~A.}
  \bibnamefont{Chekhonin}}, \bibinfo{journal}{Opt.\ Spectrosk.}
  \textbf{\bibinfo{volume}{93}}, \bibinfo{pages}{1033}
  (\bibinfo{year}{2002}{\natexlab{a}})
  [\bibinfo{journal}{Opt.\ Spectrosc.} \textbf{\bibinfo{volume}{93}},
  \bibinfo{pages}{955} (\bibinfo{year}{2002})].

\bibitem[{\citenamefont{Bagayev
  et~al.}(2002{\natexlab{b}})\citenamefont{Bagayev, Egorov, Mekhov, Moroshkin,
  Fedorov, Chekhonin, Davliatchine, and Kindel}}]{26}
\bibinfo{author}{\bibfnamefont{S.~N.} \bibnamefont{Bagayev}},
  \bibinfo{author}{\bibfnamefont{V.~S.} \bibnamefont{Egorov}},
  \bibinfo{author}{\bibfnamefont{I.~B.} \bibnamefont{Mekhov}},
  \bibinfo{author}{\bibfnamefont{P.~V.} \bibnamefont{Moroshkin}},
  \bibinfo{author}{\bibfnamefont{A.~N.} \bibnamefont{Fedorov}},
  \bibinfo{author}{\bibfnamefont{I.~A.} \bibnamefont{Chekhonin}},
  \bibinfo{author}{\bibfnamefont{E.~M.} \bibnamefont{Davliatchine}},
  \bibnamefont{and} \bibinfo{author}{\bibfnamefont{E.}~\bibnamefont{Kindel}},
  \bibinfo{journal}{Proc.\ SPIE} \textbf{\bibinfo{volume}{4748}},
  \bibinfo{pages}{45} (\bibinfo{year}{2002}{\natexlab{b}}).

\bibitem[{\citenamefont{Bagaev et~al.}(2003)\citenamefont{Bagaev, Egorov,
  Mekhov, Moroshkin, and Chekhonin}}]{27}
\bibinfo{author}{\bibfnamefont{S.~N.} \bibnamefont{Bagaev}},
  \bibinfo{author}{\bibfnamefont{V.~S.} \bibnamefont{Egorov}},
  \bibinfo{author}{\bibfnamefont{I.~B.} \bibnamefont{Mekhov}},
  \bibinfo{author}{\bibfnamefont{P.~V.} \bibnamefont{Moroshkin}},
  \bibnamefont{and} \bibinfo{author}{\bibfnamefont{I.~A.}
  \bibnamefont{Chekhonin}}, \bibinfo{journal}{Opt.\ Spectrosk.}
  \textbf{\bibinfo{volume}{94}}, \bibinfo{pages}{99} (\bibinfo{year}{2003})
  [\bibinfo{journal}{Opt.\ Spectrosc.} \textbf{\bibinfo{volume}{94}},
  \bibinfo{pages}{92} (\bibinfo{year}{2003})].

\bibitem[{\citenamefont{Allen and Eberly}(1975)}]{28}
\bibinfo{author}{\bibfnamefont{L.}~\bibnamefont{Allen}} \bibnamefont{and}
  \bibinfo{author}{\bibfnamefont{J.~H.} \bibnamefont{Eberly}},
  \emph{\bibinfo{title}{Optical Resonance and Two-level Atoms}}
  (\bibinfo{publisher}{Wiley}, \bibinfo{address}{New York},
  \bibinfo{year}{1975}).

\bibitem[{\citenamefont{Lamb}(1971)}]{29}
\bibinfo{author}{\bibfnamefont{G.~L.} \bibnamefont{Lamb}},
  \bibinfo{journal}{Rev.\ Mod.\ Phys.} \textbf{\bibinfo{volume}{43}},
  \bibinfo{pages}{99} (\bibinfo{year}{1971}).

\bibitem[{\citenamefont{Crisp}(1970)}]{30}
\bibinfo{author}{\bibfnamefont{M.~D.} \bibnamefont{Crisp}},
  \bibinfo{journal}{Phys.\ Rev.\ A} \textbf{\bibinfo{volume}{1}},
  \bibinfo{pages}{1604} (\bibinfo{year}{1970}).

\bibitem[{\citenamefont{Diels and Hahn}(1973)}]{31}
\bibinfo{author}{\bibfnamefont{J.~C.} \bibnamefont{Diels}} \bibnamefont{and}
  \bibinfo{author}{\bibfnamefont{E.~L.} \bibnamefont{Hahn}},
  \bibinfo{journal}{Phys.\ Rev.\ A} \textbf{\bibinfo{volume}{8}},
  \bibinfo{pages}{1084} (\bibinfo{year}{1973}).

\bibitem[{\citenamefont{Miklaszewski}(1995)}]{32}
\bibinfo{author}{\bibfnamefont{W.}~\bibnamefont{Miklaszewski}},
  \bibinfo{journal}{J.\ Opt.\ Soc.\ Am.\ B} \textbf{\bibinfo{volume}{12}},
  \bibinfo{pages}{1909} (\bibinfo{year}{1995}).

\bibitem[{\citenamefont{Mattar and Newstein}(1977)}]{33}
\bibinfo{author}{\bibfnamefont{F.~P.} \bibnamefont{Mattar}} \bibnamefont{and}
  \bibinfo{author}{\bibfnamefont{M.~C.} \bibnamefont{Newstein}},
  \bibinfo{journal}{IEEE J.\ Quantum Electron.} \textbf{\bibinfo{volume}{13}},
  \bibinfo{pages}{507} (\bibinfo{year}{1977}).

\bibitem[{\citenamefont{Gibbs et~al.}(1977)\citenamefont{Gibbs, Boelger,
  Mattar, Newstein, Forster, and Toschek}}]{34}
\bibinfo{author}{\bibfnamefont{H.~M.} \bibnamefont{Gibbs}},
  \bibinfo{author}{\bibfnamefont{B.}~\bibnamefont{Boelger}},
  \bibinfo{author}{\bibfnamefont{F.~P.} \bibnamefont{Mattar}},
  \bibinfo{author}{\bibfnamefont{M.~C.} \bibnamefont{Newstein}},
  \bibinfo{author}{\bibfnamefont{G.}~\bibnamefont{Forster}}, \bibnamefont{and}
  \bibinfo{author}{\bibfnamefont{P.~E.} \bibnamefont{Toschek}},
  \bibinfo{journal}{Phys.\ Rev.\ Lett.} \textbf{\bibinfo{volume}{37}},
  \bibinfo{pages}{1743} (\bibinfo{year}{1977}).

\bibitem[{\citenamefont{Gawlik et~al.}(2001)\citenamefont{Gawlik, Shuker,
  and Gallagher}}]{35}
\bibinfo{author}{\bibfnamefont{W.} \bibnamefont{Gawlik}},
  \bibinfo{author}{\bibfnamefont{R.}~\bibnamefont{Shuker}},
  \bibnamefont{and} \bibinfo{author}{\bibfnamefont{A.} \bibnamefont{Gallagher}},
  \bibinfo{journal}{Phys.\ Rev.\ A} \textbf{\bibinfo{volume}{64}},
  \bibinfo{pages}{021801(R)} (\bibinfo{year}{2001}).

\bibitem[{\citenamefont{Crenshaw and Cantrell}(1989)}]{36}
\bibinfo{author}{\bibfnamefont{M.~E.} \bibnamefont{Crenshaw}}, \bibnamefont{and}
  \bibinfo{author}{\bibfnamefont{C.~D.} \bibnamefont{Cantrell}},
  \bibinfo{journal}{Phys.\ Rev.\ A} \textbf{\bibinfo{volume}{39}},
  \bibinfo{pages}{126} (\bibinfo{year}{1989}).

\bibitem[{\citenamefont{Starostin et~al.}(1977)\citenamefont{Starostin, Panteleev,
  Lebedev, Rotin, Leonov, and Chekhov}}]{37}
\bibinfo{author}{\bibfnamefont{A.~N.} \bibnamefont{Starostin}},
  \bibinfo{author}{\bibfnamefont{A.~A.}~\bibnamefont{Panteleev}},
  \bibinfo{author}{\bibfnamefont{V.~I.} \bibnamefont{Lebedev}},
  \bibinfo{author}{\bibfnamefont{S.~V.} \bibnamefont{Rotin}},
  \bibinfo{author}{\bibfnamefont{A.~G.}~\bibnamefont{Leonov}}, \bibnamefont{and}
  \bibinfo{author}{\bibfnamefont{D.~I.} \bibnamefont{Chekhov}},
  \bibinfo{journal}{Zh.\ Eksp.\ Teor.\ Fiz.} \textbf{\bibinfo{volume}{108}},
  \bibinfo{pages}{1203} (\bibinfo{year}{1995}) [\bibinfo{journal}{Sov.\ Phys.\
  JETP} \textbf{\bibinfo{volume}{81}},
  \bibinfo{pages}{660} (\bibinfo{year}{1995})].

\bibitem[{\citenamefont{Egorov et~al.}()\citenamefont{Egorov, Lebedev, Mekhov,
  Moroshkin, Chekhonin, and Bagayev}}]{38}
\bibinfo{author}{\bibfnamefont{V.~S.} \bibnamefont{Egorov}},
  \bibinfo{author}{\bibfnamefont{V.~N.} \bibnamefont{Lebedev}},
  \bibinfo{author}{\bibfnamefont{I.~B.} \bibnamefont{Mekhov}},
  \bibinfo{author}{\bibfnamefont{P.~V.} \bibnamefont{Moroshkin}},
  \bibinfo{author}{\bibfnamefont{I.~A.} \bibnamefont{Chekhonin}},
  \bibnamefont{and} \bibinfo{author}{\bibfnamefont{S.~N.}
  \bibnamefont{Bagayev}}, \eprint{quant-ph/0308155}.

\end{thebibliography}
\end{document}